\documentclass[aps,prl,twocolumn,superscriptaddress,showpacs,preprintnumbers,amsmath,amssymb]{revtex4}
\usepackage{epsfig}
\usepackage{graphicx}
\graphicspath{{ps}}

{\renewcommand{\thefootnote}{\fnsymbol{footnote}}

\def\myspecial#1{\special{#1}}       
\def\mySpecialText{PRD RAFT v.7  2006/12/1  Contact: chwang@nuu.edu.tw}
\myspecial{!userdict begin /bop-hook{gsave 60 100 translate 90 rotate
    /Times-BoldItalic findfont 18 scalefont setfont
    0 0 moveto 0.70 setgray
    (\mySpecialText)
    show grestore}def end}                                                      
\def\anti#1#2{\vbox{\ialign{##\crcr
    \hrulefill$\smash{\phantom{\scriptstyle#2}}$\crcr 
    \noalign{\kern-1pt\nointerlineskip\vskip 0.25ex}  
    $\hfil{#1}^{#2}\hfil$\crcr}}}                     

\begin{document}

\preprint{\vbox{ \hbox{ }
                 \hbox{ }
                 \hbox{ }
                 \hbox{ }
                 \hbox{ }
                 \hbox{ }
}}

\title{ \quad\\[0.5cm] Measurement of charmless $B$ Decays to $\eta K^*$ 
and $\eta \rho$ }
%
\affiliation{Budker Institute of Nuclear Physics, Novosibirsk}
\affiliation{Chiba University, Chiba}
\affiliation{University of Cincinnati, Cincinnati, Ohio 45221}
\affiliation{Department of Physics, Fu Jen Catholic University, Taipei}
\affiliation{Justus-Liebig-Universit\"at Gie\ss{}en, Gie\ss{}en}
\affiliation{The Graduate University for Advanced Studies, Hayama, Japan}
\affiliation{Gyeongsang National University, Chinju}
\affiliation{Hanyang University, Seoul}
\affiliation{University of Hawaii, Honolulu, Hawaii 96822}
\affiliation{High Energy Accelerator Research Organization (KEK), Tsukuba}
\affiliation{Hiroshima Institute of Technology, Hiroshima}
\affiliation{University of Illinois at Urbana-Champaign, Urbana, Illinois 61801}
\affiliation{Institute of High Energy Physics, Chinese Academy of Sciences, Beijing}
\affiliation{Institute of High Energy Physics, Vienna}
\affiliation{Institute of High Energy Physics, Protvino}
\affiliation{Institute for Theoretical and Experimental Physics, Moscow}
\affiliation{J. Stefan Institute, Ljubljana}
\affiliation{Kanagawa University, Yokohama}
\affiliation{Korea University, Seoul}
\affiliation{Kyoto University, Kyoto}
\affiliation{Kyungpook National University, Taegu}
\affiliation{Swiss Federal Institute of Technology of Lausanne, EPFL, Lausanne}
\affiliation{University of Ljubljana, Ljubljana}
\affiliation{University of Maribor, Maribor}
\affiliation{University of Melbourne, Victoria}
\affiliation{Nagoya University, Nagoya}
\affiliation{Nara Women's University, Nara}
\affiliation{National Central University, Chung-li}
\affiliation{National United University, Miao Li}
\affiliation{Department of Physics, National Taiwan University, Taipei}
\affiliation{H. Niewodniczanski Institute of Nuclear Physics, Krakow}
\affiliation{Nippon Dental University, Niigata}
\affiliation{Niigata University, Niigata}
\affiliation{University of Nova Gorica, Nova Gorica}
\affiliation{Osaka City University, Osaka}
\affiliation{Osaka University, Osaka}
\affiliation{Panjab University, Chandigarh}
\affiliation{Peking University, Beijing}
\affiliation{University of Pittsburgh, Pittsburgh, Pennsylvania 15260}
\affiliation{Princeton University, Princeton, New Jersey 08544}
\affiliation{RIKEN BNL Research Center, Upton, New York 11973}
\affiliation{Saga University, Saga}
\affiliation{University of Science and Technology of China, Hefei}
\affiliation{Seoul National University, Seoul}
\affiliation{Shinshu University, Nagano}
\affiliation{Sungkyunkwan University, Suwon}
\affiliation{University of Sydney, Sydney NSW}
\affiliation{Tata Institute of Fundamental Research, Bombay}
\affiliation{Toho University, Funabashi}
\affiliation{Tohoku Gakuin University, Tagajo}
\affiliation{Tohoku University, Sendai}
\affiliation{Department of Physics, University of Tokyo, Tokyo}
\affiliation{Tokyo Institute of Technology, Tokyo}
\affiliation{Tokyo Metropolitan University, Tokyo}
\affiliation{Tokyo University of Agriculture and Technology, Tokyo}
\affiliation{Toyama National College of Maritime Technology, Toyama}
\affiliation{University of Tsukuba, Tsukuba}
\affiliation{Virginia Polytechnic Institute and State University, Blacksburg, Virginia 24061}
\affiliation{Yonsei University, Seoul}
  \author{C.~H.~Wang}\affiliation{National United University, Miao Li} 
  \author{K.~Abe}\affiliation{High Energy Accelerator Research Organization (KEK), Tsukuba} 
  \author{I.~Adachi}\affiliation{High Energy Accelerator Research Organization (KEK), Tsukuba} 
  \author{H.~Aihara}\affiliation{Department of Physics, University of Tokyo, Tokyo} 
  \author{D.~Anipko}\affiliation{Budker Institute of Nuclear Physics, Novosibirsk} 
  \author{T.~Aushev}\affiliation{Swiss Federal Institute of Technology of Lausanne, EPFL, Lausanne}\affiliation{Institute for Theoretical and Experimental Physics, Moscow} 
  \author{A.~M.~Bakich}\affiliation{University of Sydney, Sydney NSW} 
  \author{E.~Barberio}\affiliation{University of Melbourne, Victoria} 
  \author{A.~Bay}\affiliation{Swiss Federal Institute of Technology of Lausanne, EPFL, Lausanne} 
  \author{K.~Belous}\affiliation{Institute of High Energy Physics, Protvino} 
% ...
\collaboration{The Belle Collaboration}

\noaffiliation

\begin{abstract}
We report measurements of branching fractions
and $CP$ asymmetries for 
$B \to \eta K^*$ and $B \to \eta \rho$ decays.
These results are obtained from a $414\,{\rm
fb}^{-1}$ data sample collected at   
the $\Upsilon(4S)$ resonance
with the Belle detector at the KEKB 
asymmetric-energy $e^+ e^-$ collider.
We measure the following branching fractions:
${\mathcal B}(B^0 \to \eta K^{*0})  =
(15.2 \pm 1.2 \pm 1.0)\times 10^{-6}$ and
${\mathcal B}(B^+\to\eta K^{*+}) = 
(19.3^{+2.0}_{-1.9}\pm 1.5)\times 10^{-6}$,
where the first error is statistical and the second systematic.
We also find a $2.7\sigma$ excess in the
$B^+\to\eta \rho^+$ mode and measure
$\mathcal B(B^+\to\eta \rho^+)=
(4.1^{+1.4}_{-1.3}\pm 0.4)\times 10^{-6} <6.5 \times 10^{-6}$
at 90\% confidence level.
For $B^0 \to \eta \rho^0$ decays, we determine the upper limit
$\mathcal B(B^0\to\eta \rho^0) < 1.9 \times 10^{-6}$
at 90\% confidence level.
The partial rate asymmetries are
${\mathcal A_{CP}}(\eta K^{*0}) = 0.17 \pm 0.08 \pm 0.01$, 
${\mathcal A_{CP}}(\eta K^{*+}) = 0.03 \pm 0.10 \pm 0.01$, and 
${\mathcal A_{CP}}(\eta \rho^+) = -0.04^{+0.34}_{-0.32}\pm 0.01$.

\end{abstract}
\maketitle

\tighten
{\renewcommand{\thefootnote}{\fnsymbol{footnote}}}
\setcounter{footnote}{0}
\section{Introduction} 
\begin{figure}[htb]
\includegraphics[width=0.48\textwidth]{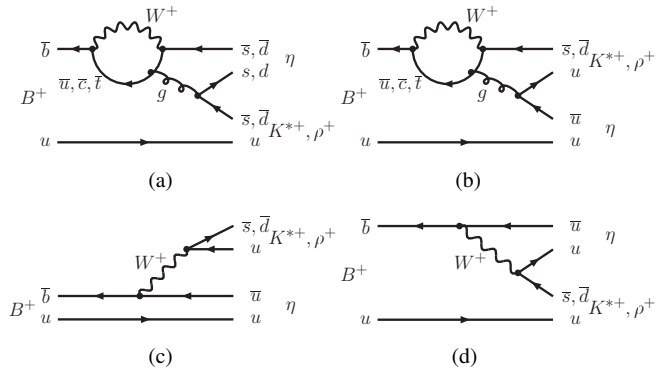}
\caption{Feynman diagrams for $B^+ \to \eta K^{*+}$
and $\eta \rho^+$ decays. The corresponding neutral decays are 
similar except that the spectator quark
becomes a $d$ and (b) and (c) diagrams do not exist.}
\label{feynman}
\end{figure}

Charmless hadronic $B$ decays play an important role in 
understanding $CP$ violation in the $B$ meson system.
The decays $B \to \eta K^*$ and $B \to \eta \rho$
are key examples.
In the standard model (SM), penguin (tree) diagrams are
expected to dominate in $B \to \eta K^*$ ($B \to \eta \rho$) decays (Fig.~\ref{feynman}).
The large branching fraction for $B \to \eta K^*$ compared to that
for $B \to\eta K$~\cite{CLEO,BELLEETA,BABARETA}
can be explained qualitatively in terms of the interference between 
non-strange and strange components of the $\eta$ meson, but is higher 
than recent theoretical predictions~\cite{ali,cheng,pqcd,rosner}.
In a similar vein, the larger measured branching fraction for 
charged ($B^+ \to \eta K^{*+}$) versus neutral ($B^0 \to \eta K^{*0}$) 
decays may suggest an additional SU(3)-singlet 
contribution~\cite{pqcd,rosner,hou} or constructive interference 
between SM penguin and tree amplitudes or 
between SM and new physics penguin amplitudes.
Throughout this paper, the inclusion of charge-conjugate 
modes is implied unless stated otherwise.%

In the standard model,
direct $CP$ violation (DCPV) occurs in decays
due to interference between
two (or more) amplitudes that have different strong and 
weak phases.  
The partial rate asymmetry can be written as
\begin{eqnarray}
{\mathcal A}_{CP}(B\to f) = {\Gamma({\overline B \to \bar f})
- \Gamma({B \to f}) \over \Gamma({\overline B \to \bar f})
+ \Gamma({B \to f})}  \nonumber \\
  = { 2 |M_1| |M_2| \sin\Delta\delta \sin\Delta\phi \over
  	|M_1|^2 + |M_2|^2 + 2 |M_1| |M_2| \cos\Delta\delta \cos\Delta\phi}, 
\end{eqnarray}
where $\overline B$ and $\bar f$ are the $CP$-conjugate 
states, and $\Delta\delta$ ($\Delta\phi$) is the difference 
of the strong (weak) phases between amplitudes $M_1$ 
and $M_2$. Here, the amplitude $M_1$($M_2$) represents the
sum of the amplitudes from penguin (tree) diagrams having a
common weak phase.
The asymmetry will be sizable when the two type of amplitudes are of
comparable strength with significant phase differences.
However, $B \to \eta K^*$ and $B \to \eta \rho$ decay rates 
are expected to be dominated by penguin diagrams for $\eta K^*$
and tree diagrams for $\eta \rho$, 
so ${\mathcal A}_{CP}$ is expected to be small.
On the other hand, amplitudes arising from new physics may 
interfere with these SM amplitudes to generate a sizable 
${\mathcal A}_{CP}$ value.
The experimental results~\cite{BABARETA} suggest that DCPV is small,
albeit with large statistical uncertainties. 
\section{Data Set and Apparatus}  
This analysis is based on a data sample collected at the 
$\Upsilon$(4S) resonance with the Belle detector~\cite{BelleNIM} 
at the KEKB~\cite{kekb} accelerator.
The data sample corresponds to an integrated 
luminosity of $414\,{\rm fb}^{-1}$ and contains 
$449 \times 10^{6}$ $B \overline{B}$ pairs. 

The Belle detector is designed to measure charged particles and
photons with high efficiency and precision. 
Charged particle tracking is provided
by a silicon vertex detector (SVD) and a central drift chamber
(CDC) that surround the interaction region. 
The charged particle
acceptance covers the laboratory polar angle between $\theta=17^\circ$
and $150^\circ$, measured from the $z$ axis that is aligned 
anti-parallel to the positron beam. 
Charged hadrons are distinguished by combining the
responses from an array of silica aerogel Cherenkov counters
(ACC), a barrel-like array of 128 time-of-flight scintillation
counters (TOF), and $dE/dx$ measurements in the CDC. The combined
response provides $K/\pi$ separation of at least $2.5\sigma$ for
laboratory momentum up to 3.5~GeV/$c$. Electromagnetic showers are
detected in an array of 8736 CsI(Tl) crystals (ECL) located inside
the magnetic volume, which covers the same solid angle as the
charged particle tracking system. The 1.5-T magnetic field
is contained via a flux return that consists 
of 4.7 cm thick steel plates, interleaved with resistive plate 
counters used for tracking muons.
Two inner detector configurations were used. A 2.0 cm beampipe and 
a 3-layer silicon vertex detector were used for the first sample of
$152 \times 10^{6}$ $B \overline{B}$ pairs, while a 1.5 cm beampipe, 
a 4-layer silicon detector and a small-cell inner drift chamber
were used to record the remaining $297 \times 10^{6}$ $B \overline{B}$ 
pairs~\cite{SVD2}.

We calculate the acceptance and study backgrounds using 
Monte Carlo (MC) simulation.
For these simulation studies, 
the signal events, generic $b \to c$ decays and 
charmless rare $B$ decays
are generated with the EVTGEN~\cite{evtgen} event generator.
The continuum MC
events are generated with the $e^+e^- \to \gamma^* \to q \bar{q}$ 
process in the JETSET~\cite{jetset} generator. 
The GEANT3~\cite{geant} package is used for detector simulation.
\section{Event Selection and Reconstruction}
Hadronic events are selected based on the charged track multiplicity
and total visible energy sum, which give an efficiency
greater than 99\% for $B\overline B$ events.
All primary charged tracks are required to be consistent with coming from
the run-dependent 
interaction point within $\pm 2\,{\rm cm}$ along the $z$ axis 
and within $\pm 1.5\, {\rm cm}$ in the transverse plane.
Particle identification (PID) is based on the likelihoods 
${\mathcal L_{\rm K}}$ and 
${\mathcal L_\pi}$ for charged kaons and pions, respectively. 
These likelihoods are calculated from CDC, TOF, and ACC 
information.
A higher value of ${\mathcal L_{\rm K}}$/(${\mathcal L_\pi+ \mathcal L_{\rm K}}$)  
indicates a more kaon-like particle. 
In this analysis, PID cuts are applied to all charged particles 
except those associated with 
$K^0_S \to \pi^+ \pi^-$ decays. 
Unless explicitly specified, the PID cuts are 
${\mathcal L_{\rm K}}$/(${\mathcal L_\pi+\mathcal L_{\rm K}}$) $> 0.6$ for kaons
and $< 0.4$ for pions. The corresponding efficiencies are $85$\% for
kaons and $89$\% for pions; $8$\% of pions are misidentified
as kaons and $11$\% of kaons are misidentified as pions.

We form $\pi^0$ candidates from photon pairs with an invariant mass 
between 118 MeV/$c^2$ and $150\, {\rm MeV}/c^2$ ($2.5\sigma$).
The photon energies must exceed 50 MeV, and the $\pi^0$ momentum in 
the center-of-mass (CM) frame must exceed 0.35 GeV/$c$.
$K_S^0$ candidates are reconstructed from pairs 
of oppositely charged tracks whose invariant mass lies 
within $\pm 10\, {\rm MeV}/c^2$ ($2.5\sigma$) of the $K_S^0$ meson mass.
We also require that the vertex of the $K^0_S$  
be well-reconstructed and displaced from the interaction point, 
and that the $K^0_S$ momentum direction be consistent 
with the $K^0_S$ flight direction.  
\subsection{$\eta$ Meson Reconstruction}
Candidate $\eta$ mesons are reconstructed in the
$\eta \to \gamma \gamma$ and $\eta \to \pi^+\pi^- \pi^0$ modes.
If one of the photons from the 
former $\eta$ decay mode can be paired with another photon and have a 
reconstructed $\gamma\gamma$ mass within $3\sigma$ of the $\pi^0$ meson 
mass, the $\eta$ candidate is discarded.
We relax the PID requirement 
for charged pions from the latter $\eta$ decay mode to 
${\mathcal L_{\rm K}}$/(${\mathcal L_\pi+L_{\rm K}}$) $< 0.9$.
Candidate $\eta$ mesons are required to satisfy
the following mass selections:
$500$~MeV/$c^2$$\le M_{\gamma\gamma} \le 575$~MeV/$c^2$
and $537$~MeV/$c^2$$\le M_{\pi^+\pi^-\pi^0} \le 557$~MeV/$c^2$,
where the reconstructed mass resolutions are 12~MeV/$c^2$ for 
$\eta \to \gamma \gamma$ and 3.5~MeV/$c^2$ for 
$\eta \to \pi^+ \pi^- \pi^0$.  
When reconstructing the $B$ meson candidate, 
the momentum of the $\eta$ candidate is 
recalculated by applying the $\eta$ mass constraint. 
The $\eta \to \gamma\gamma$ candidates must satisfy $|\cos\theta^*| 
< 0.90$, where $\theta^*$ is the angle between the photon 
direction and the direction
of the CM frame in the $\eta$ rest frame;
this requirement suppresses 
soft photon combinatorial and $B \to K^* \gamma$ feed-across backgrounds. 
\subsection{$K^*$ and $\rho$ Meson Reconstruction}
$K^{*0}$ candidates are reconstructed from $K^+\pi^-$ and 
$K_S^0\pi^0$ pairs, while the $K^{*+}$ candidates are reconstructed from $K^+\pi^0$ 
and $K_S^0\pi^+$ pairs.
These candidates are required to have  
reconstructed masses within $\pm$75~MeV/$c^2$ of the nominal value~\cite{PDG}.
Candidate $\rho^0$ ($\rho^+$) mesons are reconstructed from 
$\pi^-\pi^+$ ($\pi^0\pi^+$) pairs. Each combination is required to have 
a reconstructed mass within $\pm$150 MeV/$c^2$ of the nominal value~\cite{PDG}.   
\subsection{$B$ Meson Reconstruction}
The $B$ meson candidates are reconstructed from $\eta K^{*0}$, 
$\eta K^{*+}$, $\eta \rho^0$, and $\eta \rho^+$ combinations. They are 
characterized by the beam-energy-constrained mass 
$M_{\rm bc}$ = $\sqrt{E^2_{\rm beam}/c^4-|P_B/c|^2}$ and the energy difference 
$\Delta E = E_B - E_{\rm beam}$, where $E_{\rm beam} = 5.29$~GeV, and 
$P_B$ and $E_B$ are the momentum and energy, respectively, of the $B$ 
candidate in the CM frame. 
We define the fit region in the $M_{\rm bc}$--$\Delta E$ plane as 
$M_{\rm bc} > 5.2\,{\rm GeV}/c^2$ and $|\Delta E| < 0.25\,{\rm GeV}$. We 
define the signal region as the overlap of the bands 
$M_{\rm bc} > 5.27\,{\rm GeV}/c^2$ and $|\Delta E| < 0.05\,{\rm GeV}$.

From signal MC, we find that $8-10$\% of the events contain 
multiple $B$ candidates. 
Only one $B$ candidate per event is 
retained for the likelihood fit.
If there are multiple $\eta$ candidates, we choose the one 
with the smallest $\chi^2$ of the fit with a mass (vertex and mass) constraint 
to the kinematics of the $\eta$ meson in the case of 
$\eta \to \gamma \gamma$ ($\eta \to \pi^+\pi^-\pi^0$) decays.
Among the $B$ candidates made of the same $\eta$ candidate, we choose
the one with the smallest vertex $\chi^2$ in the cases
of $K^{*0} \to K^+ \pi^-$ or $\rho^0 \to \pi^+ \pi^-$; 
or then the one with the mass closest to
nominal in the case of $K^{*+}\to K^0\pi^+$;
or then the $\pi^0$ mass closest to the nominal in all other cases. 
\section{Background Suppression}
The dominant background for exclusive two-body
$B$ decays comes from the 
$e^+e^- \to \gamma^* \to q \bar{q}$ continuum ($q = u,\,d,\,s,\,c$),
which has a jet-like event topology in contrast to more spherical
$B\overline{B}$ events. The other major backgrounds involve feed-across
from these and other charmless $B$ decays. 
The background from $b \to c$ decays has
a small impact because the $M_{\rm bc}$ and $\Delta E$ 
distributions do not peak in the signal region.
In this analysis, the fit does not distinguish non-resonant 
$B \to \eta K \pi$ decays from $B \to \eta K^*$ decays, since 
they have the same $M_{\rm bc}$ and $\Delta E$ distributions. 
The non-resonant contribution is estimated and subtracted 
independently using the $K\pi$ invariant mass distributions 
of the fitted $B$-decay yields.
\subsection{Continuum Background}
Signal and continuum events are distinguished in two steps. Here, all the
variables are calculated in the CM frame. 
First, we require $|\cos\theta_T| < 0.9$, where $\theta_T$ is defined as the 
angle between the $\eta$ direction of a $B$ candidate and the thrust axis 
from all particles in the event not associated with that $B$ 
candidate. This retains 90\% of signal and removes $\sim$56\% of 
continuum. Second, a likelihood ${\cal L}_s$ 
(${\cal L}_c$) for signal (continuum) is formed from two independent 
variables---$\cos\theta_B$, where $\theta_B$ is the polar angle of the 
$B$ candidate momentum direction, and a Fisher discriminant~\cite{fisher} 
${\cal F} = \vec{\alpha}\cdot \vec{R}$ that combines seven 
event shape variables: $\cos\theta_T$, 
$S_{\perp}$(the sum of the magnitudes of the momenta transverse to 
the $\eta$ direction for all particles more than $45^{\circ}$ away from 
the $\eta$ axis, divided by the sum of the magnitudes of the momenta
of all particles not from the candidate $B$ meson~\cite{sperp}), 
and the five modified Fox-Wolfram 
moments~\cite{fw} $R_2^{so}$, $R_4^{so}$, $R_2^{oo}$, $R_3^{oo}$, and 
$R_4^{oo}$. The Fisher discriminant's weight vector $\vec{\alpha}$ is 
determined to maximize the separation between signal events and
continuum background using MC data; these Fox-Wolfram moments 
are used since they are not correlated with $M_{\rm bc}$. 
The likelihood ratio 
${\cal R} = {\cal L}_s / ({\cal L}_s + {\cal L}_c)$, which peaks near one
for signal and near zero for continuum,  
is used to distinguish signal from continuum.

The distribution of ${\cal R}$ is found to depend somewhat on the 
event's $B$ flavor tagging quality parameter $r$~\cite{btag}, which ranges 
from zero for no flavor identification to unity for unambiguous flavor 
assignment. We partition the data into three $r$ regions, $r\le 0.5$,
$0.5 < r \le 0.75$, and $r > 0.75 $.  
In each $r$ region, the optimal cut on ${\cal R}$ is determined 
by maximizing the significance 
$N_S/\sqrt{N_S+N_B}$, where $N_S$ and $N_B$ are the retained number of signal 
and continuum background events selected in MC samples. For cut optimization
studies, we assume the branching fractions of $2.0\times 10^{-5}$ for $\eta K^*$,
$5.0\times 10^{-6}$ for $\eta \rho^+$, 
and $1.0\times 10^{-6}$ for $\eta \rho^0$.
For $B^0 \to \eta K^{*0}, \eta \to \gamma\gamma, K^{*0} \to K^+ \pi^-$ decays, 
a typical cut of ${\mathcal R} > 0.4$ is $\sim 87$\% efficient for signal 
and removes $\sim 67$\% of the continuum background for data in the region 
$r \le 0.5$, and a cut of ${\mathcal R} > 0.2$ 
is $\sim 94$\% efficient for signal 
and removes $\sim 53$\% of the continuum background 
for data in the region $r > 0.75 $.

\subsection{Backgrounds from $B$ Decays}
$B \to K^*(\rho) \gamma$ is the dominant charmless
$B$-decay background for $B \to \eta K^*(\rho)$,
$\eta \to \gamma\gamma$ decays. 
The $\eta \to \gamma\gamma$ selection, $|\cos\theta^*| 
< 0.90$, removes 85\% of this background.
%
%
To further suppress it, 
we pair each photon from the $\eta \to \gamma\gamma$ candidate
with the $K^*$ or $\rho$ candidate and reject those events 
where $M_{\rm bc}>5.27\,{\rm GeV}/c^2$ and 
$-0.2\,{\rm GeV} < \Delta E < 0.1\,{\rm GeV}$.
We thus remove 96\% of this background and
retain 93\% of the signal events.
For $B \to \eta\rho$ decays, a measurable contribution
from other charmless $B$ and $b \to c$ decays 
remains (see Table ~\ref{etarho_bg}).
The contributions of these backgrounds are taken into account in the analysis.
\begin{table}[hb]
\begin{center}
\caption{Estimated $B-decay$ background contributions
in the fit region to $B \to \eta \rho$ from $b \to c$ ($N_{\rm bc}$), 
charmless $B$ decay ($N_{\rm r}$), and
$\eta K^*$ feed-across ($N_{\rm feed}$) and 
measured yields from all sources ($N$) 
but dominated by residual continuum background,
after application of the $\cos\theta_T$ 
and ${\mathcal R}$ cuts.
$N_{\rm bc}$ and $N_{\rm r}$ are estimated from MC samples.}
\begin{tabular}{lccc|c} \\ \hline \hline
Mode & $N_{\rm bc}$ & $N_{\rm r}$ & $N_{\rm feed}$ &$N$ \cr\hline
$\eta_{\gamma\gamma} \rho^0$ & $62$ & $81$ & $17$ & $2931$  \cr
$\eta_{\pi\pi\pi^0} \rho^0$ & $67$ & $27$ & $5$ & $1063$  \cr \hline
$\eta_{\gamma\gamma} \rho^+$ & $148$ & $74$ & $3$ & $4169$  \cr
$\eta_{\pi\pi\pi^0} \rho^+$ & $76$ & $22$ & $1$ & $1809$  \cr \hline
\hline
\end{tabular}
\label{etarho_bg}
\end{center}
\end{table}
\section{Analysis Procedure}
Signal yields are obtained using an extended unbinned
maximum likelihood fit to the $M_{\rm bc}$ and $\Delta E$
distributions (2-D ML) for events that satisfy the 
$\cos\theta_T$ and ${\cal R}$ requirements.

For $N$ input candidates, the likelihood is defined as
\begin{eqnarray}
 L &&= \frac{e^{-(N_S+N_{qq}+N_{\rm bc}+N_{\rm r}+N_{\rm feed})}}{N!}
\prod_{i=1}^{N}(N_{S} P_{S_i} +  \nonumber \\ 
&& N_{qq} P_{{qq}_i} + N_{\rm bc} P_{{\rm bc}_i} 
+ N_{\rm r} P_{{\rm r}_i} + N_{\rm feed} P_{{\rm feed}_i}),
\end{eqnarray}
where $P_{S_i}$, $P_{{qq}_i}$, $P_{{\rm bc}_i}$,
$P_{{\rm r}_i}$ and $P_{{\rm feed}_i}$ are the probability 
density functions for event $i$, with measured values
$M_{{\rm bc},i}$ and $\Delta E_i$, to arise from signal, continuum
background, $b \to c$ background, charmless $B$ decay background, and
feed-across background, respectively.
%
%
The small yields $N_{\rm bc}$, $N_{\rm r}$, and $N_{\rm feed}$ are 
fixed from the MC analysis.

The continuum, $ b \to c$ and charmless $B$-decay
background $\Delta E$ probability density functions (PDF) are modeled by 
second- or third-order polynomial functions. The continuum and $b \to c$
background components in $M_{\rm bc}$ are
modeled by a smooth function~\cite{argus}. 
To account for the peaking behavior of $M_{\rm bc}$
in the signal region from charmless $B$ decay
backgrounds, we use the sum of two bifurcated-Gaussian functions
to model the distributions. 
The bifurcated Gaussian combines the left half of a wide-resolution 
Gaussian with the right half of a narrow-resolution Gaussian, both having 
a common mean.
For $B \to \eta \rho$ decays, the $M_{\rm bc}$ and $\Delta E$
distributions from $\eta K^*$ feed-across will behave like signal 
with a $\Delta E$ 
shift of $-50$ MeV. The PDF shape for each contribution
is determined from MC. 
The first-order coefficient of the 
continuum-background $\Delta E$ polynomial and the 
parameters of the $M_{\rm bc}$ function are allowed to float in each fit.

For the signal $\Delta E$ distribution, we combine 
two bifurcated-Gaussian functions. 
The first accounts for $60$-$80$\% of the total area and 
the wider second models the low-energy tail.
$M_{\rm bc}$ is weakly correlated with $\Delta E$, so we construct 
separate bifurcated Gaussians for $M_{\rm bc}$ in the three ranges 
$|\Delta E| < 0.05\,{\rm GeV}$, $0.05\,{\rm GeV} < |\Delta E| < 
0.1\,{\rm GeV}$, and $0.1\,{\rm GeV} < |\Delta E| < 0.25\,{\rm GeV}$.
The parameters of these functions are estimated from MC first, then calibrated
with a large control sample 
of $B^+ \to \anti{D}{0} \pi^+$, $ \anti{D}{0} \to K^+\pi^-\pi^0$ decays.

For decays with more than one sub-decay process,
the final results are obtained
by fitting the sub-decay modes simultaneously  
with the expected 
efficiencies folded in and with the branching fraction
as the common output.
%
%
The statistical significance ($\Sigma$) of the signal
is defined as $\sqrt{-2\ln(L_0/L_{\rm max})}$,
where $L_0$ and $L_{\rm max}$ denote
the likelihood values for zero signal events
and the best fit numbers, respectively.

The 90\% confidence level (C.L.) upper limit $x_{90}$ on the signal yield
is calculated from the equation
\begin{eqnarray*}
\frac{\int_0^{x_{90}} L(x) \,dx}{\int_0^\infty L(x) \,dx}
 = 90\% \,.
\end{eqnarray*}
To incorporate the systematic uncertainty in the calculation of $x_{90}$,
the likelihood function is smeared with a Gaussian function 
with the resolution from the systematic uncertainty. That smeared likelihood function
is also used to calculate the significance of the signal including the systematic uncertainty.
\section{Measurements of Branching Fractions}
\subsection{Efficiencies and Corrections}
The overall reconstruction efficiency $\epsilon$
is first obtained using MC samples
and then multiplied by PID efficiency corrections
obtained from data.
The PID efficiency correction is
determined using $D^{*+} \to D^0 \pi^+$,
$D^0 \to K^- \pi^+$ data samples.
Other MC efficiency corrections are determined
by comparing data and MC predictions
for other well-known processes.
The charged-particle tracking efficiency correction is studied using
a high-momentum $\eta$ sample and is determined by comparing 
the ratios of $\eta \to \pi^+ \pi^- \pi^0$ to $\eta \to \gamma\gamma$ 
in data and MC. 
The same high-momentum $\eta$ sample is also used for $\pi^0$ 
reconstruction efficiency 
corrections by comparing the ratio of
$\eta \to \pi^0 \pi^0 \pi^0$ to $\eta \to \gamma\gamma$ between
the data and MC sample.
The $K^0_S$ reconstruction efficiency is verified by
comparing four $K^*(892)$ decay channels
($K^+\pi^-$, $K^+\pi^0$, $K^0_S\pi^+$, $K^0_S\pi^0$) in
inclusive $K^*$ and exclusive $B \to J/\psi K^*$ samples.
The ${\mathcal R}$ cut efficiency correction is determined
using $B^+ \to \anti{D}{0} \pi^+$ decays.
For $\eta$ and $K^*$ reconstruction and mass cuts,
we use the high-momentum $\eta$ and $K^*$ sample for
the efficiency correction studies.
The above studies show good agreement 
between data and MC; 
the reconstruction and selection efficiencies 
differ by about $2$\%.
The PID, $\pi^0$, $\eta$ and $K^*$
reconstruction efficiency corrections are applied and
the systematic uncertainties are also obtained from the above studies.
\begin{table}[htb]
\caption{Summary of results for each channel listed in the first column.
The measured signal yield ($N_S$),
reconstruction efficiency ($\epsilon$), 
total efficiency ($\epsilon_{\rm {tot}}$) including the 
secondary branching fraction,
statistical significance ($\Sigma$) and
measured branching fractions are shown.
Uncertainties shown in second and sixth columns are statistical only.
For the final combined branching fractions,
corrections for contributions from non-resonant
or higher resonance components have been applied.
The total systematic uncertainties are given,
and the combined significances include the systematic uncertainties.
\label{result}}
\vspace{0.1cm}
\begin{tabular}{lccccc}
\hline \hline
Mode & $N_S$ & $\epsilon$(\%) &
$\epsilon_{\rm {tot}}$(\%) & $\Sigma$ & ${\mathcal B}(10^{-6})$\\
\hline\hline
$\eta_{\gamma\gamma} K^{*0}_{K^+\pi^-}$  &  $336.2^{+30.1}_{-29.2}$
              & $16.9$ & $4.4$ & $14.2$ &
	      $16.9\pm 1.5$ \\ 
$\eta_{\pi\pi\pi^0} K^{*0}_{K^+\pi^-}$  & $93.4^{+14.6}_{-13.8}$
              & $9.8$ & $1.5$  & $8.7$ &
               $14.1^{+2.2}_{-2.1}$ \\
$\eta_{\gamma\gamma} K^{*0}_{K^0\pi^0}$  & $20.1^{+7.5}_{-6.7}$
              & $2.1$ & $0.27$ & $3.6$ &
	      $16.7^{+6.3}_{-5.6}$ \\ 
$\eta_{\pi\pi\pi^0} K^{*0}_{K^0\pi^0}$  & $9.5^{+5.0}_{-4.2}$
              & $1.3$ & $0.098$  & 2.6 &
               $21.6^{+11.5}_{-9.7}$ \\\hline
$\eta K^{*0}$  & -
              & - & -  & $15.7$ &
               {$15.2\pm 1.2 \pm 1.0$} \\\hline\hline
$\eta_{\gamma\gamma} K^{*+}_{K^+ \pi^0}$ &  $79.8^{+16.1}_{-15.3}$
              & $6.7$ & $0.88$ & 6.1 &
	      $20.1^{+4.1}_{-3.9}$ \\ 
$\eta_{\pi\pi\pi^0} K^{*+}_{K^+ \pi^0}$ & $24.1^{+8.7}_{-7.9}$
              & $4.2$ & $0.32$ & $3.5$ &
	      $16.9^{+6.1}_{-5.6}$ \\ 
$\eta_{\gamma\gamma} K^{*+}_{K^0 \pi^+}$ & $120.3^{+16.2}_{-15.4}$
              & $4.5$ & $1.2$ & $10.1$ &
	      $22.6^{+3.1}_{-2.9}$ \\ 
$\eta_{\pi\pi\pi^0} K^{*+}_{K^0 \pi^+}$ & $29.2^{+7.3}_{-6.6}$
              & $2.6$ & $0.38$ & $6.2$ &
	      $16.9^{+4.3}_{-3.8}$ \\ \hline
$\eta K^{*+}$ & -
              & - & - & $12.3$ &
	      {$19.3^{+2.0}_{-1.9}\pm 1.5$} \\ \hline\hline
$\eta_{\gamma\gamma} \rho^0$  & $19.5^{+11.3}_{-10.4}$
              & $8.9$ & $3.5$ & $2.1$ &
	      $1.25^{+0.73}_{-0.67}$ \\ 
$\eta_{\pi\pi\pi^0} \rho^0$  & $0.9^{+4.6}_{-3.9}$
              & $5.5$ & $1.2$  & $0.2$ &
               $0.17^{+0.84}_{-0.66}$ \\ \hline
$\eta \rho^0$  & -
              & - & -  & $1.3$ &
               $0.84^{+0.56}_{-0.51}\pm 0.19$ \\ \hline\hline
$\eta_{\gamma\gamma} \rho^{+}$ &$38.1^{+16.1}_{-15.2}$
              & $5.5$ & $2.2$ & $2.6$ &
	      $3.9^{+1.7}_{-1.6}$\\ 
$\eta_{\pi\pi\pi^0} \rho^+ $ & $15.8^{+8.9}_{-8.0}$
              & $3.50$ & $0.79$ & 2.1 &
	       $4.4^{+2.5}_{-2.2}$  \\ \hline
$\eta \rho^+$ & -
              & - & - & $2.7$ &
	      {$4.1^{+1.4}_{-1.3}\pm 0.4$} \\ \hline\hline
\end{tabular}
\end{table}
\subsection{Fit Results}
\begin{figure}[htb]
\includegraphics[width=0.53\textwidth]{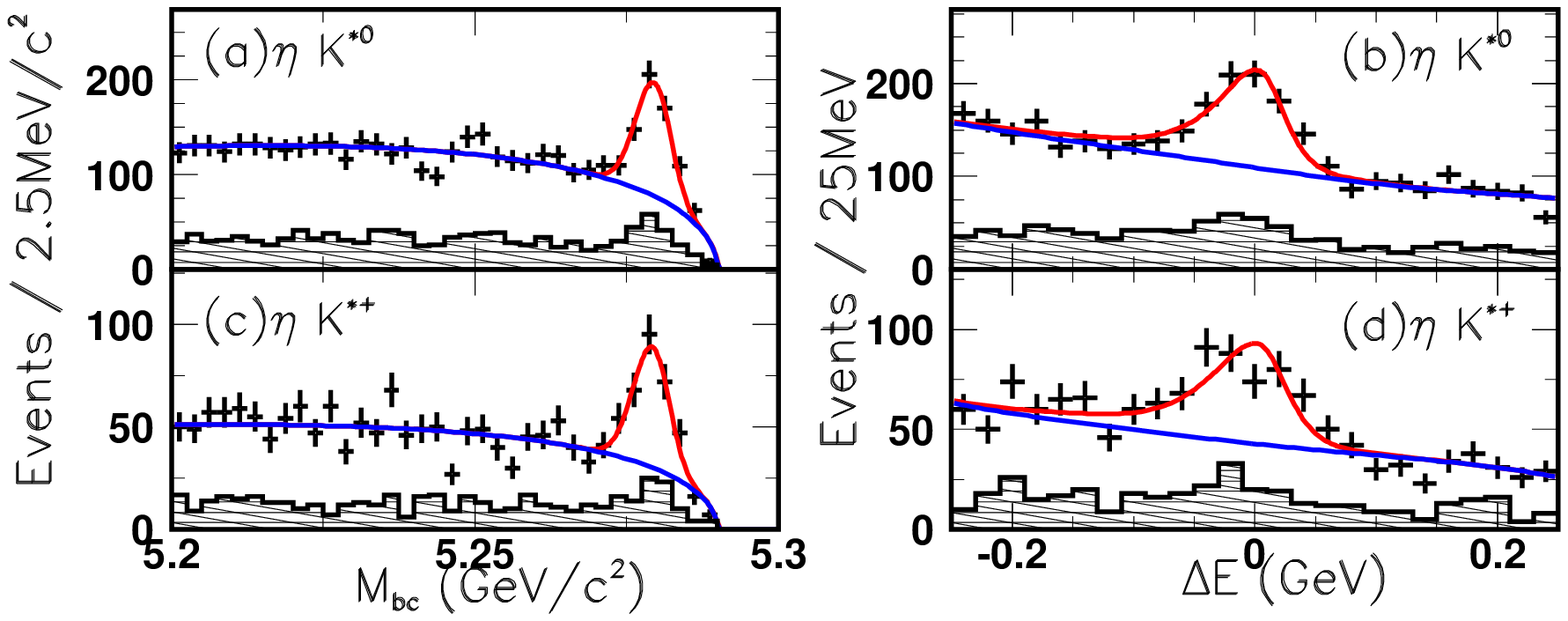}
\caption{Projections on $M_{\rm bc}$ (for the signal slice in $\Delta E$)
and $\Delta E$ (for the signal slice in $M_{\rm bc}$)
for $\eta K^{*0}$ (a,b) and $\eta K^{*+}$ (c,d) with the 
expected signal and background
curves overlaid. The shaded area represents 
$\eta \to \pi^+ \pi^- \pi^0$ decays.}
\label{etafigmb}
\end{figure}
\begin{figure}[htb]
\includegraphics[width=0.53\textwidth]{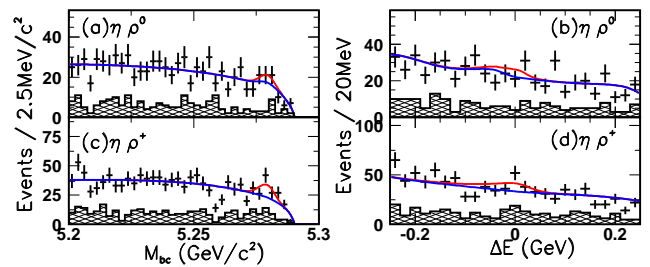}
\caption{Projections on $M_{bc}$ (for the signal slice in $\Delta E$)
and $\Delta E$ (for the signal slice in $M_{\rm bc}$)
from 2-D ML fit results
for $\eta \rho^{0}$ (a,b) and $\eta \rho^{+}$ (c,d) with 
the expected signal and background
curves overlaid. The shaded area represents 
$\eta \to \pi^+\pi^-\pi^0$ decays.}
\label{etarho}
\end{figure}

The fitted signal yields and branching fractions are shown in 
Table~\ref{result}.
Several consistency checks are made, including
tighter ${\mathcal R}$ cuts as well as 1-D ML $M_{\rm bc}$
and $\Delta E$ fits,
and they are all found to be consistent.
The total observed yields
are $N_{\eta K^{*0}} = 459.2^{+34.6}_{-33.3}$ for
$B^0 \to \eta K^{*0}$,
$N_{\eta K^{*+}} = 253.4^{+25.5}_{-24.0}$
for $B^+ \to \eta K^{*+}$,
$N_{\eta \rho^0} = 20.4^{+12.2}_{-11.0}$ for
$B^0 \to \eta \rho^0$ and
$N_{\eta \rho^+} = 53.9^{+18.4}_{-17.1}$
for $B^+ \to \eta \rho^+$.
%
Figure \ref{etafigmb} shows the projections
of the data and the fits onto $M_{\rm bc}$ (for events in $\Delta E$ 
signal slice) and $\Delta E$ (for events in the $M_{\rm bc}$ signal slice) 
for the $B \to \eta K^*$ decays, while Fig.~\ref{etarho}
shows the corresponding projections for the $B \to \eta \rho$ decays. 
\subsection{Non $K^*(892)$ Components}
\begin{figure}[htb]
\includegraphics[width=0.51\textwidth]{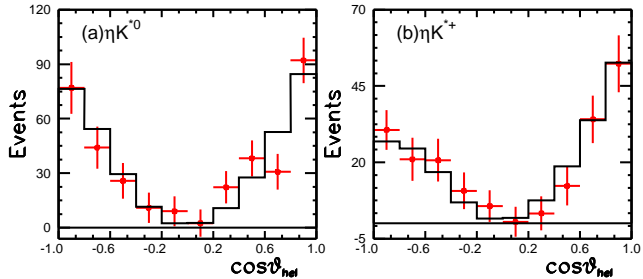}
\caption{Distributions of the $K^*$ helicity for the (a) $K^{*0}$
and (b) $K^{*+}$ modes in case of $B \to \eta K^*$
decays. The overlaid histograms represent the distributions 
from MC normalized 
by the 2-D fit results. 
}
\label{ksthe}
\end{figure}

The background-subtracted $K^*$ helicity distributions within 
the $M_{\rm bc}$ and $\Delta E$ signal regions (Fig.~\ref{ksthe})
are consistent with the expectation from $\eta K^*$ 
decays, indicating no significant S-wave or higher
resonance contribution in the $K^*$ mass region.
The $K^{*0}(K^{*+})$ helicity angle ($\theta_{\rm {hel}}$) is the angle
between the $\pi^-(\pi^0,K^0)$ direction and the opposite
of the $B$ direction in the $K^*$ rest frame.
The binned $\chi^2$ per degree of freedom is 
$\chi^2/N = 9.7/10$ for $K^{*0}$ and
$\chi^2/N = 3.6/10$ for $K^{*+}$.
\begin{figure}[htb]
\includegraphics[width=0.51\textwidth]{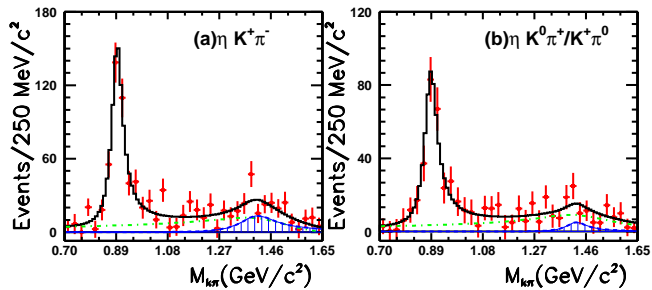}
\caption{Fitted yields vs. the $K\pi$ invariant mass for the (a) $(K\pi)^0$
and (b) $(K\pi)^+$ modes. The overlaid functions are the results of
a binned $\chi^2$ fit. 
The dashed line represents the contribution
from the D-wave $K_2^*(1430)$, and
the dot-dash line represents the LASS S-wave parameterization. 
LASS parameterization parameters, widths of P-wave and 
D-wave functions are allowed to float in the fits.}
\label{kpifit}
\end{figure}
\begin{figure}[hb]
\includegraphics[width=0.51\textwidth]{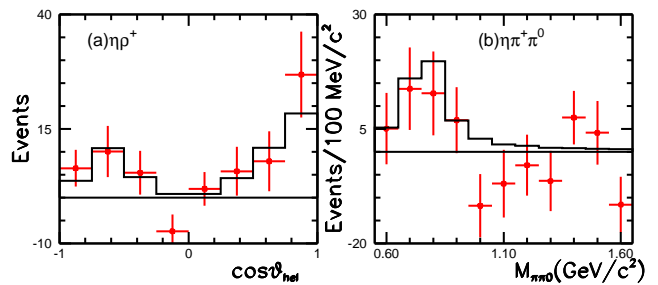}
\caption{Fitted yields vs. (a) $\rho^+$ helicity and
(b)$\pi^+\pi^0$ invariant mass from $B\to\eta \rho^+$ decays. 
The overlaid histograms are expected distributions from MC and
normalized by the 2-D fit results with $\chi^2/N = 6.8/8$ for
(a) and $\chi^2/N = 13.6/11$ for (b).}
\label{etarhohelfig}
\end{figure}

We use a 2D-ML fit to the $K\pi$ invariant mass distributions to 
evaluate the small S-wave or higher 
resonance $K\pi$ contaminations in the $K^*$ 
mass region.
A clear excess in the higher $K\pi$ invariant
mass region is observed (Fig.~\ref{kpifit}).
To estimate these contributions,
we fit the distributions with a P-wave relativistic 
Breit-Wigner function for the $K^*$(892), 
a D-wave relativistic Breit-Wigner function for the $K_2^*(1430)$
resonance and an ad hoc function for the S-wave contribution.
Several functions are used to model the
S-wave contribution in the $K\pi$ mass region
$1.0$~GeV/$c^2$ to $1.5$~GeV/$c^2$, including
a $K_0^{*}(1350)$ resonance\cite{lass}, the LASS
distribution\cite{lass}, and a threshold function.
The LASS distribution contains a non-resonant S-wave background function 
interfering with an S-wave $K_0^*(1430)$ resonance.
%
%
Figure~\ref{kpifit} shows one example of our fitting results 
with the S-wave contribution modeled by the LASS distribution.
Based on these studies with various S-wave functions, 
the non-resonant $K\pi$ contributions
are ($5.6 \pm 3.0$)\% for $\eta K^{*0}$ and 
($5.0 \pm 3.0$)\% for $\eta K^{*+}$ decays. 
These corrections are applied to
the final branching fraction measurements of $B \to \eta K^*$.

For $B^+ \to \eta \rho^+$ decays, 
we examine the properties of the $\rho^+$ candidates through
2-D ML fits in bins of $\pi^+\pi^0$ invariant mass 
and $\rho^+$ helicity.
Although statistically limited,
a $\rho^+$ mass peak and a
polarized $\rho^+$ helicity distribution are 
observed (Fig.~\ref{etarhohelfig}) 
and are consistent with the expectation from $\eta \rho^{+}$.
Due to the limited statistics for 
$B^+ \to \eta \rho^+$ decays, 
a larger systematic error for the non-resonant or higher resonance
contributions is assigned with no corrections applied.

\section{Systematic Error}
Systematic errors, enumerated in Table~\ref{system}, arise from
efficiency corrections, non-resonant corrections and fitting.
The main sources of uncertainties in the efficiency 
corrections are from
the reconstruction of low-momentum charged tracks,
low-energy photon finding, and the ${\mathcal R}$ cut efficiency.
The systematic errors include contributions of
$1$\% for ${\mathcal R}$ cuts, $1$\% per reconstructed charged 
particle, $0.5$\%
for each charged particle identification, 
$4$\% for $\pi^0$ reconstruction, $4.5$\% for $K^0_S$ reconstruction,
and $2$\% for $\eta$ reconstruction with $\eta \to \gamma \gamma$.
We use $B^+ \to \anti{D}{0} \pi^+$ decays
to estimate the uncertainties in the signal
PDF's used for fitting $M_{\rm bc}$ and $\Delta E$ 
by comparing the mean and the width of the $M_{\rm bc}$ and
$\Delta E$ distributions between the $B^+ \to \anti{D}{0} \pi^+$ 
data and the MC sample.
In Table~\ref{system}, ``Fit'' means the systematic uncertainty from
the PDF function modeling. ``$B_s$'' means the systematic uncertainty
from the braching fractions of $\eta$ and $K^*(\rho)$ decays,
which is obtained from the PDG tables~\cite{PDG}. 
``Non-resonant'' means the systematic uncertainty from the 
non-resonant or higher resonant contributions 
in the $K^*(\rho)$ mass window regions, which is
obtained from the studies of a 2D-ML fit to the
$K\pi$ invariant mass distribution.
For MC estimated $b \to c$ and charmless $B$ decay 
backgrounds (``$N_{\rm bc}$'' and ``$N_{\rm r}$'', respectively), 
we vary the estimated yields by $\pm 50$\% and  
refit the data. 
The difference between the resulting signal yield and 
the nominal value is taken as an additional systematic 
error.
The overall relative systematic errors are
$6.5$\% for $\eta K^{*0}$, $7.5$\% for $\eta K^{*+}$,
$22.1$\% for $\eta \rho^0$ 
and $9.6$\% for $\eta \rho^+$. 
\begin{table}[htb]
\caption{Relative systematic errors for $\eta K^*$ 
and $\eta \rho$. 
The unit is in percent $(\%)$.}
\begin{tabular}{lcccc} \hline\hline
Contribution & $\eta K^{*0}$ & $\eta K^{*+}$ & $\eta \rho^0$ &
$\eta \rho^+$ \\\hline
charged track/$K^0_S$ reconstruction   & 2.9  & 4.4 & 3.3 & 1.8 \\
$\pi^0$/$\eta \to \gamma\gamma$ selection & 2.8  & 3.5 & 3.2 & 4.8 \\
$\eta$ mass window      & 2.0  & 2.0 & 2.0 & 2.0 \\
$K^*(\rho)$ mass window  & 2.0  & 2.0 & 2.0 & 2.0 \\
PID correction                & 1.3  & 1.0 & 1.6 & 0.8 \\
$\mathcal {R}$ requirement                   & 1.0  & 1.0 & 1.0 & 1.0 \\
Fit                           & $1.7$  & $1.8$ & $7.5$ & $4.2$ \\
$N_{\rm bc}$, $N_{\rm r}$     & $1.2$  & $1.2$ & $7.5$ & $0.4$ \\
$\eta K^*$ feed-across & - & - & $17.4$ & $1.1$ \\
$N_{B\overline B}$ & 1 & 1 & 1 & 1 \\
$B_s$        & 1.0 & 1.0 & 1.4 & 0.8 \\
Non-resonant        & 3.0 & 3.0 & 6.0 & 6.0 \\\hline 
Total      & $6.5$ & $7.5$ & $22.1$ & $9.6$ \\\hline\hline
\end{tabular}
\label{system}
\end{table}

\section{$\mathcal A_{CP}$ Measurements}
We measure $\mathcal A_{CP}$ for
$B \to \eta K^*$ and $B^+ \to \eta \rho^+$.
To account for the wrong-tag fraction $w$, the true value of $\mathcal A_{CP}$
is related to the measured $\mathcal A^{\rm obs}_{CP}$ via
${\mathcal A^{\rm obs}_{CP}} = (1-2w){\mathcal A_{CP}}$.
Among the decay modes we study, only those in which the $\mathcal A_{CP}$ values
are determined by low momentum charged pions
have a significant $w$: the wrong-tag fractions 
for $K^{*+} \to K^0\pi^+$ is $\sim 1.5$\%
for $\eta K^{*+}$ decays and $\sim 2.0$\% 
for $\eta \rho^+$ decays, while
other decays have $w < 0.1$\%. Since the result for
$\eta K^{*+}$ is obtained from a simultaneous fit to all four sub-decay modes
with roughly equal statistics for $K^{*+} \to K^+ \pi^0$
and $K^{*+} \to K^0 \pi^+$, 
the wrong-tag effect for $\mathcal A_{CP}$
is less than 0.7\%. 
Therefore, the only mode where we apply a correction due to the wrong
tag fraction is $B^+ \to \eta \rho^+$, where we estimate $w=2$\%.
To incorporate the $CP$ asymmetry in the fit, the coefficients of the 
signal and continuum background PDF's in the likelihood are modified as follows: 
$N_S \to {1\over 2}N_S(1 - q{\cal A}_{CP}^{\rm obs})$ and 
$N_{qq} \to {1\over 2}N_{qq}(1 - q{\cal A}_{CP,qq})$,
where $q = +1 (-1)$ for a $B(\overline B)$ meson tag and 
$\mathcal A^{\rm obs}_{CP}$,$\mathcal A_{CP,qq}$ are the
$\mathcal A_{CP}$ outputs for signal and 
continuum, respectively.
The results are ${\mathcal A^{\rm obs}_{CP}}(\eta K^{*0})=0.17 \pm 0.08$,
${\mathcal A^{\rm obs}_{CP}}(\eta K^{*+})=0.03 \pm 0.10$ and
${\mathcal A^{\rm obs}_{CP}}(\eta \rho^+)=-0.04^{+0.34}_{-0.32}$.

Since the systematic errors in the reconstruction of the $\eta$ candidates and
the number of $B\overline{B}$ events cancel in the ratio,
the systematic uncertainty on $\mathcal {A_{CP}}$
comes mainly from the charge asymmetry in the identification of charged kaons 
and the fitting PDF's. To estimated the fitting-PDF systematic uncertainty,
we apply the same procedures as in the branching fraction measurements.
The relative systematic errors from fitting PDF's are estimated to be
$3$\% for $\eta K^{*0}$, $13$\% for $\eta K^{*+}$,
and $27$\% for $\eta \rho^+$. 
The efficiency asymmetry for the PID of charged kaons is $0.01$
in absolute value.
\section{Summary}
In summary, we report measurements of the exclusive two-body 
charmless hadronic $B \to \eta K^*$ and $B \to \eta \rho$ decays 
with high statistics.
Our results 
are consistent with previous measurements~\cite{CLEO,BABARETA}
and confirm that the branching fractions for
$B^0\to \eta K^{*0}$ and $B^+\to \eta K^{*+}$ are large.
The branching fractions obtained are
${\mathcal B}(B^0\to\eta K^{*0}) =
(15.2 \pm 1.2 \pm 1.0)\times 10^{-6}$, and
${\mathcal B}(B^+\to\eta K^{*+}) =
(19.3^{+2.0}_{-1.9}\pm 1.5)\times 10^{-6}$,
where the first error is statistical and the second systematic.
Our measurements indicate that the branching fraction for
$B^+ \to \eta K^{*+}$ is $1.4 \sigma$ higher than that for 
$B^0 \to \eta K^{*0}$.
A $2.7\sigma$ excess is seen for $B^+ \to \eta \rho^+$ decays.
The branching fraction and 90\% C.L. upper limits
for $B \to \eta \rho$ decays are
$\mathcal B$($B^+\to\eta \rho^+$)=
$(4.1^{+1.4}_{-1.3}\pm 0.4)\times 10^{-6}$($< 6.5 \times 10^{-6}$) and
$\mathcal B$($B^0\to\eta \rho^0$)$< 1.9 \times 10^{-6}$. 
The measurements of the $B \to \eta \rho$ branching fractions are consistent 
with theoretical predictions~\cite{ali,cheng,pqcd,rosner}.

We have measured the direct $CP$ asymmetry in the $B \to \eta K^*$ and 
$B^+ \to \eta \rho^+$ channels.
Our results are
${\mathcal A_{CP}}(\eta K^{*0}) = 0.17 \pm 0.08 \pm 0.01,
{\mathcal A_{CP}}(\eta K^{*+}) =  0.03 \pm 0.10 \pm 0.01$, and
${\mathcal A_{CP}}(\eta \rho^+) = -0.04^{+0.34}_{-0.32}\pm 0.01$,
all consistent with no asymmetry.
\section{Acknowledgments} 
We thank the KEKB group for the excellent operation of the
accelerator, the KEK cryogenics group for the efficient
operation of the solenoid, and the KEK computer group and
the National Institute of Informatics for valuable computing
and Super-SINET network support. We acknowledge support from
the Ministry of Education, Culture, Sports, Science, and
Technology of Japan and the Japan Society for the Promotion
of Science; the Australian Research Council and the
Australian Department of Education, Science and Training;
the National Science Foundation of China and the Knowledge
Innovation Program of the Chinese Academy of Sciences under
contract No.~10575109 and IHEP-U-503; the Department of
Science and Technology of India; 
the BK21 program of the Ministry of Education of Korea, 
the CHEP SRC program and Basic Research program 
(grant No.~R01-2005-000-10089-0) of the Korea Science and
Engineering Foundation, and the Pure Basic Research Group 
program of the Korea Research Foundation; 
the Polish State Committee for Scientific Research; 
the Ministry of Education and Science of the Russian
Federation and the Russian Federal Agency for Atomic Energy;
the Slovenian Research Agency;  the Swiss
National Science Foundation; the National Science Council
and the Ministry of Education of Taiwan; and the U.S.\
Department of Energy.

\end{document}